\documentclass[journal]{IEEEtran}
\usepackage{titletoc}
\usepackage{diagbox}
\usepackage{algorithmic}
\usepackage{makecell}
\usepackage{algorithm}
\usepackage{array}
\usepackage{amsmath,amssymb,amsfonts}
\usepackage[caption=false,font=normalsize,labelfont=sf,textfont=sf]{subfig}
\usepackage{cite}
\usepackage{textcomp}
\usepackage{stfloats}
\usepackage{url}
\usepackage{verbatim}
\usepackage{graphicx}
\usepackage{lipsum}
\usepackage{xcolor}
\usepackage{makecell}
\usepackage{multirow}
\usepackage{amsmath}
\usepackage{color}
\usepackage[colorlinks,
            linkcolor=blue,
            anchorcolor=blue,
            citecolor=blue,
            urlcolor=blue
            ]{hyperref}

\begin{document}

\title{Large Language Models based Multi-Agent Framework for Objective Oriented Control Design in Power Electronics  \\

\thanks{}
}

\author{
        \vskip 1em
                Chenggang Cui, Jiaming Liu, Junkang Feng,  Peifeng Hui, Amer M. Y. M. Ghias, and Chuanlin Zhang
       
}
\maketitle

\definecolor{limegreen}{rgb}{0.2, 0.8, 0.2}
\definecolor{forestgreen}{rgb}{0.13, 0.55, 0.13}
\definecolor{greenhtml}{rgb}{0.0, 0.5, 0.0}

\begin{abstract}
Power electronics, a critical component in modern power systems, face several challenges in control design, including model uncertainties, and lengthy and costly design cycles. This paper is aiming to propose a Large Language Models (LLMs) based multi-agent framework for objective-oriented control design in power electronics. The framework leverages the reasoning capabilities of LLMs and a multi-agent workflow to develop an efficient and autonomous controller design process. The LLM agent is able to understand and respond to high-level instructions in natural language, adapting its behavior based on the task's specific requirements and constraints from a practical implementation point of view. This novel and efficient approach promises a more flexible and adaptable controller design process in power electronics that will largely facilitate the practitioners.
\end{abstract}

\begin{IEEEkeywords}
objective oriented control design, power electronics, microgrids, large language models
\end{IEEEkeywords}

\section{Introduction}

Power electronics play a vital role in modern power systems, serving as a critical link between the source of power and the end user\cite{b1}. These systems are essential in controlling the conversion and distribution of electrical power, ensuring that power is delivered efficiently and effectively\cite{b2}. With the advent of renewable energy sources and the push for more sustainable practices, the importance of power electronics is more pronounced. However, the control methods used in power electronics present several challenges, such as he complexity of the system structure, the uncertainty of system parameters, and the variability of operating conditions\cite{b3}. 

The most common control methods used in power electronics are model-based approaches. These methods generally rely on mathematical models of the system to design effective controllers. While these methods have been widely applied into practices, they often suffer from certain limitations due to inevitable model variations, uncertainties and unmodelled dynamics \cite{b4}. Such methods may lead to inefficiencies in the design process. Moreover, the traditional controller design process is often an iterative trial-and-error process. Such process involves a significant amount of modeling, tuning, testing, and debugging iterations, which can be time-consuming and require a substantial amount of expertise and resources. This often leads to lengthy and costly design cycles for practitioners.

Recently, model-free control methods have become a trend for power electronics as reported in the literature, which bypass the demand for explicit modeling and tuning, and virtual impedance control schemes\cite{b6}. While model-free control methods in power electronics have shown promises in addressing the challenges of traditional controller design, however, they are also facing some critical issues\cite{b7}.. One of the main challenges is the implementation of the controller and safety constraints. Model-free control methods are designed without the explicit modeling and tuning process, focusing instead on achieving specific performance objectives However, the implementation of these objectives, such as reference voltage regulation in power electronics, still requires human intervention. Designers are required to define these specific performance requirement and system constraints, which can be a complex and time-consuming process. Furthermore, while model-free control methods can reduce the design cycle time, they do not completely eliminate the need for testing and debugging. These processes are still necessary to ensure the stability and reliability of the control system. In light of these challenges, there is an imperative need for further research and development in high-level control for power electronic systems, aiming to improve in areas such as objective definition, data handling, and system testing.

The objective-oriented control system (OOCS) is a novel concept that has been proposed to address the challenges associated with traditional control system design. The fundamental idea behind OOCS is to shift the design paradigm from ``How" to ``What"\cite{b12}. This means that instead of instructing the control system on how to achieve a specific objective, the designer specifies what the objective is, and the control system figures out how to achieve it. Such approach has shown several advantages compared to traditional control system design. First, it simplifies the design process. Instead of having to carefully synthesize a desirable operation condition or a profile, the designer merely needs to specify the control objective. This can significantly reduce the time and resources required for control design. In addition, by focusing on the control objective rather than the means to achieve it, the control system can explore different strategies to achieve the pre-defined  objective. This can lead to innovative solutions that may not be apparent in a traditional design process. Therefore, the objective-Oriented Control System can simplifies the design process, enhances adaptability, and can potentially improve system performance.


In this study, a Large Language Models (LLMs) based agent framework is proposed for the objective-oriented control design in the field of power electronics. The framework leverages the reasoning capabilities of the LLMs agent and a multi-agent workflow to develop an efficient power electronics controller design framework.
More specifically, the LLMs based agent can provide high-level objective-oriented control designs with only the specifies high-level objectives  and constraints (for instance, safety or resources), and the LLMs-based Agent Framework is tasked with finding the way to achieve these objectives (``how") while simultaneously satisfying all constraints. This approach allows for a more flexible and adaptable controller design process, as the LLM agent can understand and respond to high-level instructions in natural language, and can adapt its behavior based on the specific requirements and constraints of the task. This makes the LLMs-based Agent Framework a promising tool for objective-oriented control design in power electronics, providing a novel and efficient approach to solve this complex task.

The contributions of this study is as follows:

\begin{enumerate}
    \item A LLMs based multi-agent framework is proposed, specifically designed for objective-oriented control in the field of power electronics.
    \item The framework utilizes the reasoning capabilities of the LLMs agent, combined with a multi-agent workflow, to develop an efficient power electronics controller design framework.
    \item The proposed approach introduces a higher degree of flexibility and adaptability into the controller design process. The LLM agent can understand and respond to high-level instructions expressed in natural language and adjust its behavior according to the specific requirements and constraints.
\end{enumerate}

\section{Problem description and PRELIMINARIES}

\subsection{Objective-Oriented Control Design in Power Electronics}
Objective-oriented control design focuses on achieving specific objectives within power electronics systems. This approach involves setting clear objectives, planning actions to achieve those objectives, and dynamically adjusting the control strategies based on real-time feedback. Traditional methods, such as PID controllers and model predictive control, have been widely used but often lack the flexibility to adapt to rapidly changing environments.
\subsection{LLM-based Agent Frameworks}
The advent of Large Language Models (LLMs) has indeed revolutionized the design and evaluation of agent-based models
\cite{cheng2024exploring}. Recognized for their zero-shot prompting and advanced reasoning abilities, the application of LLM-based agents in controller design is a promising avenue to explore. Their ability to understand and respond to high-level instructions in natural language can simplify the controller design process, allowing designers to specify high-level objectives and constraints, with the LLM-based agent tasked with finding the best way to achieve these objectives while satisfying all constraints.

\begin{figure*}[htbp]
\centerline{\includegraphics[width=1\linewidth]{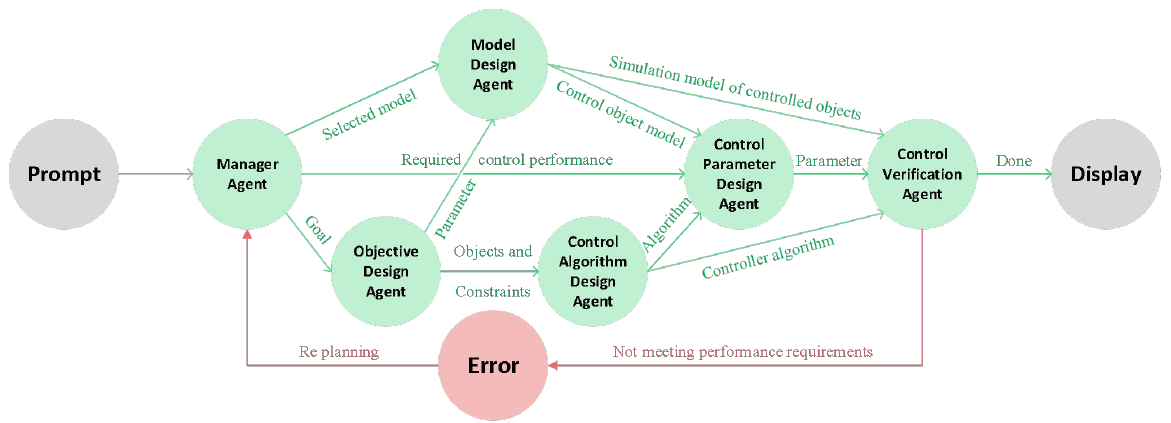}}
\caption{Workflow of Objective Oriented Controller Design.}
\label{Workflow}
\vspace{-0.7em}
\end{figure*}


\section{Objective Oriented Controller Design}

\subsection{Multi-Agent Framework Architecture}
In the design of the objective oriented Controller, we propose a multi-agent collaboration framework to decompose the objective Oriented controller design process into multiple sub-tasks, thereby achieving automation from task planning, modeling, controller design, simulation, to evaluation and verification.
\begin{figure}[htbp]
\centerline{\includegraphics[width=1\linewidth]{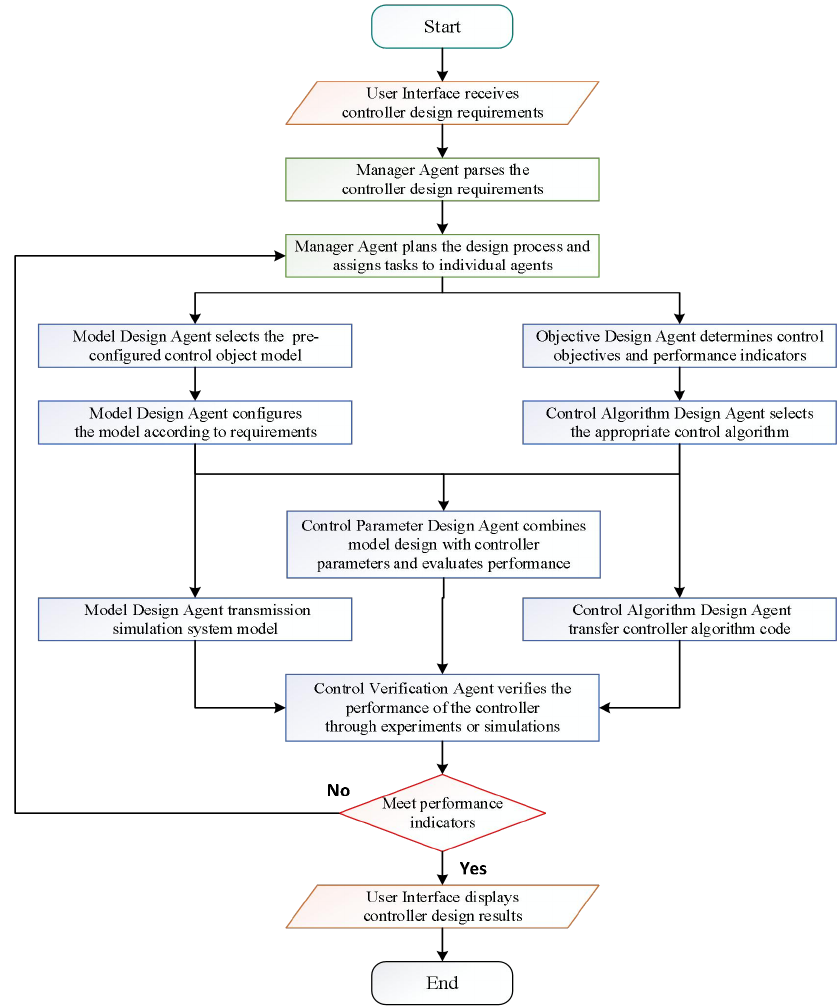}}
\caption{Flow chart of Objective Oriented Controller Design.}
\label{flowchart}
\vspace{-0.2em}
\end{figure}
The Manager Agent parses the demands proposed by the user and assigns the various tasks and requirements to the relevant functional agents. Once each agent completes their respective tasks, the controller design can be realized. This process achieves automation of the entire design process.

The basic process for objective-oriented controller design is as follows and the  workflow is shown as Figure \ref{Workflow}:

\subsubsection{Task Translation} The objective designer first translates the task into control objectives, control targets, and performance constraints for controller design.
\subsubsection{Manager} Serving as the core component of the entire system, the manager begins to invoke other agents to obtain a detailed analysis of the control objectives and targets.
\subsubsection{Agent Collaboration} These agents, including those for control objectives and control algorithms, support the invocation of some tools. For example, a simulator can use Modelica for control system simulation.
\subsubsection{Simulation and Verification} Upon receiving the replies regarding the control objectives, control algorithms, and control parameters, the manager conducts a simulation via the simulator to verify the controller design scheme, which includes controller selection and parameters.
\subsubsection{Evaluation} The evaluator is responsible for analyzing the answers in the previous attempt and providing suggestions, such as modifying the controller design scheme to meet the task requirements.

This process underscores the importance of multi-agent collaboration and iterative refinement in achieving effective and efficient controller design. It also highlights the role of simulation tools in verifying and fine-tuning the design, ensuring that the final controller design is robust and meets the specified performance criteria.This multi-agent framework does not strictly adhere to human-prescribed workflows, which allows for adaptability and flexibility in responding to unforeseen challenges. This flexibility enables the full potential of multi-agent systems.


\subsection{ Agent Role Design}

The roles of these agents are designed as follows and the flow char of objective oriented controller design is shown in Figure \ref{flowchart}.

\subsubsection{Manager Agent}
The Manager Agent is a crucial component in the multi-agent system, acting as the coordinator among the team of agents. Its primary role is to orchestrate the objective Orient Controller Design workflow, which includes identifying relevant questions, forwarding these inquiries to the appropriate agent, and summarizing the results.

\subsubsection{Objective Design Agent}

 This agent is responsible for determining the control objectives and performance indicators. It needs to set the control objectives of the system according to its actual needs, such as stabilizing the system, tracking a given trajectory, or making the system respond faster, etc. At the same time, it needs to set corresponding performance indicators, such as the stability of the system, tracking error, response time, and oscillation, etc.

\subsubsection{Model Design Agent}

The Model Design Agent is responsible for establishing a simulation model of the control object. This model can help us understand the dynamic behavior of the system and provide necessary information for the design of the controller.

\subsubsection{Control Algorithm Design Agent}
This agent is responsible for selecting the appropriate control algorithm based on the control objectives and performance indicators. For example, it can choose the classic PID controller, or it can choose more complex control methods, such as adaptive control, predictive control, model reference adaptive control, etc.

\subsubsection{Control Parameter Design Agent} 
 The Parameter Design Agent is responsible for designing the parameters of the controller. For example, for a PID controller, it needs to set the proportion, integral, and derivative coefficients. These parameters can be designed through trial and error, optimization algorithms, self-tuning algorithms, etc.
\subsubsection{Control Verification Agent} This agent is responsible for verifying the performance of the controller through experiments. If the performance of the controller does not meet expectations, it can adjust the parameters of the controller or change the control strategy based on the experimental results.

Through this multi-agent collaboration approach, we can achieve the automation and optimization of objective orient controller design. Each agent can focus on its specific task, and at the same time, through collaboration, they can collectively achieve the objective of the entire system.

\section{initial Implementation }

\begin{figure*}[htbp]
\centerline{\includegraphics[width=1\linewidth]{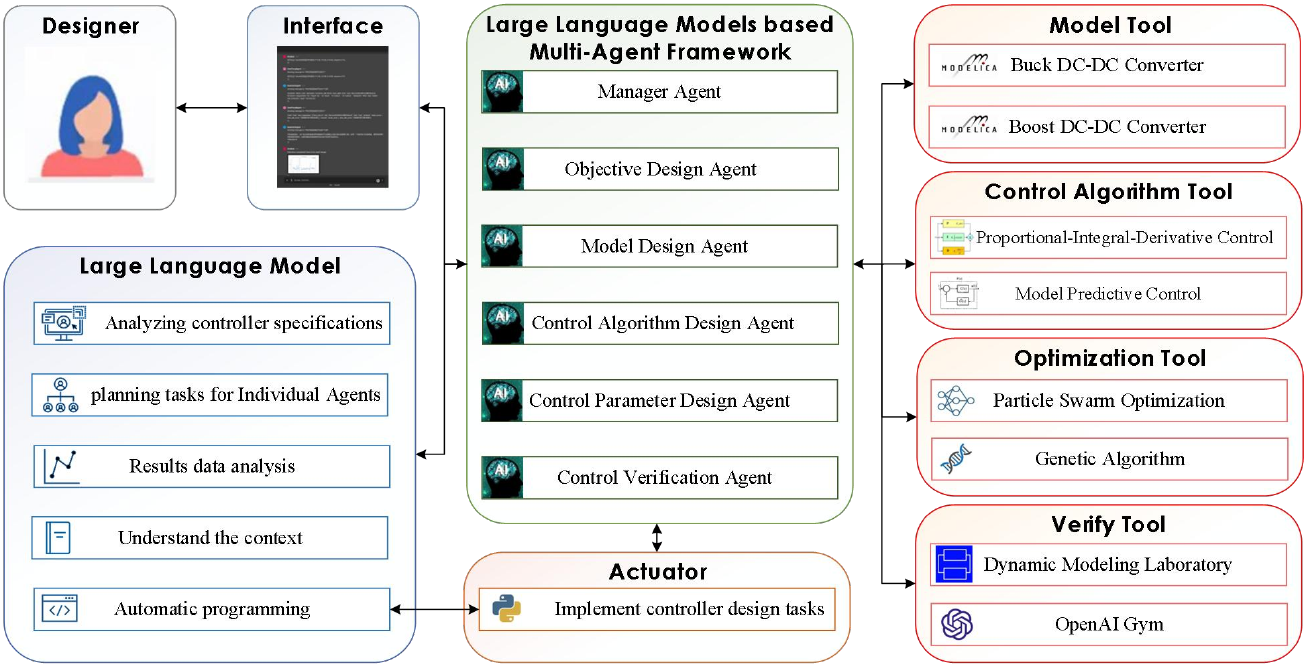}}
\caption{Initial implementation of the multi-agent framework}
\label{Design process}
\vspace{-0.6em}
\end{figure*}

Figure \ref{Design process} showcases the initial implementation of our proposed multi-agent collaboration framework for task-oriented controller design. The detailed implementation of the core agent is described as follows.

\subsection{Model Design Agent}

The primary task of the Model Design Agent is to select simulation models and generate the Parameter for power electronic devices. It receives parsed design requirements from the Manager Agent. These requirements include specific details such as the type of power electronic device, input-output voltage requirements, and load conditions. The agent then processes these requirements and selects the most appropriate Modelica model template using the Modelica model tool.

 \subsubsection{Tool} The Modelica model tool is a key tool in this process\cite{tiller2001introduction}. It may cover different types of converters, including buck, boost, and buck-boost converters. These templates serve as the starting point for model design, providing parameters that can be adjusted according to specific design requirements. Once the most suitable template is selected, the Model Design Agent adjusts the variables and parameter values within the model template.
\subsubsection{Output} Upon completion of all configurations, the agent provides the Modelica model file address and model parameters. This Modelica model file address is then passed on to the Simulation Agent. It serves as the foundation for subsequent simulation verification and parameter optimization.

\subsection{Controller Design Agent}



The primary task of the Controller Algorithm Agent is to generate controller algorithm code automatically using the Controller Algorithm Tool. This code is based on the design parameters and performance objectives provided by the Manager Agent. The Controller Algorithm Agent initiates its process by obtaining user-inputted design parameters and performance objectives from the Manager Agent.

\subsubsection{Tool}  It then employs the Controller Algorithm Tool to generate preliminary controller algorithm code, which aligns with the given design parameters and performance objectives.
The Controller Algorithm Tool empowers the Controller Algorithm Agent to generate preliminary controller algorithm code based on the design requirements, such as PID controller code. This tool references controller algorithm design templates to ensure the generated code is robust and efficient.
Agent.
\subsubsection{Output}  The output of this process is controller algorithm code that aligns with the design parameters and performance objectives. The Controller Algorithm Agent hands over the controller algorithm code to the Controller Simulation Agent to ensure the controller meets the performance.

\subsection{Controller Parameter Design Agent}

The Controller Parameter Design Agent employs intelligent optimization algorithms to optimize controller parameters, where the optimization objectives and constraints are based on parameter design provided by the Target Design Agent, guiding this optimization process.
\subsubsection{Tool}  In terms of tool usage, the Controller Parameter Design Agent selects the optimization algorithm (such as Genetic Algorithm  and Particle Swarm Optimization (PSO) algorithm \cite{marini2015particle} through the Optimization Tool and generates the final optimized parameter function.
\subsubsection{Output} The agent passes the optimized parameter function class code to the Simulation Verification Agent.

\subsection{Controller Verification Agent}

The primary task of the Controller Verification Agent is to combine model files, control algorithms, and control parameter optimization functions to perform model simulation verification through the Open AI Gym toolbox\cite{brockman2016openai}. This step is crucial in ensuring that the controller design meets the preset performance indicators.

\subsubsection{Tool} The Gym toolbox is a potent simulation tool that allows the Controller Verification Agent to simulate controller performance under actual operating conditions. The agent configures this simulation environment based on the model, controller, and control parameter information received respectively from the Model Agent, Control Algorithm Design Agent, and Control Parameter Optimization Agent. Once the simulation environment is set up, the Controller Verification Agent executes the simulation process. During this process, the agent collects performance data reflecting the controller's performance under actual operating conditions.
\subsubsection{Output} Upon completion of the simulation, the Controller Verification Agent outputs performance indicators such as steady-state error, overshoot, and response time. These indicators are then passed on to the Target Design Agent. If the simulation results do not meet the performance indicators required by the Target Design Agent, the unmet indicators are returned to the Manager Agent for task reallocation. However, if the performance indicators meet the requirements of the Target Design Agent, the simulation verification data is passed on to the Manager Agent.

\section{Demonstration On DC-DC Boost Converter}

To demonstrate the effectiveness of the proposed framework, a case study of task orient controller design is conducted for DC-DC Boost Converter. The study involves simulating a complex power system and evaluating the performance of the LLM-based agents in achieving specific control objectives.
\subsection{Controller Design Prompt}

\begin{figure}[htbp]
\centerline{\includegraphics[width=1\linewidth]{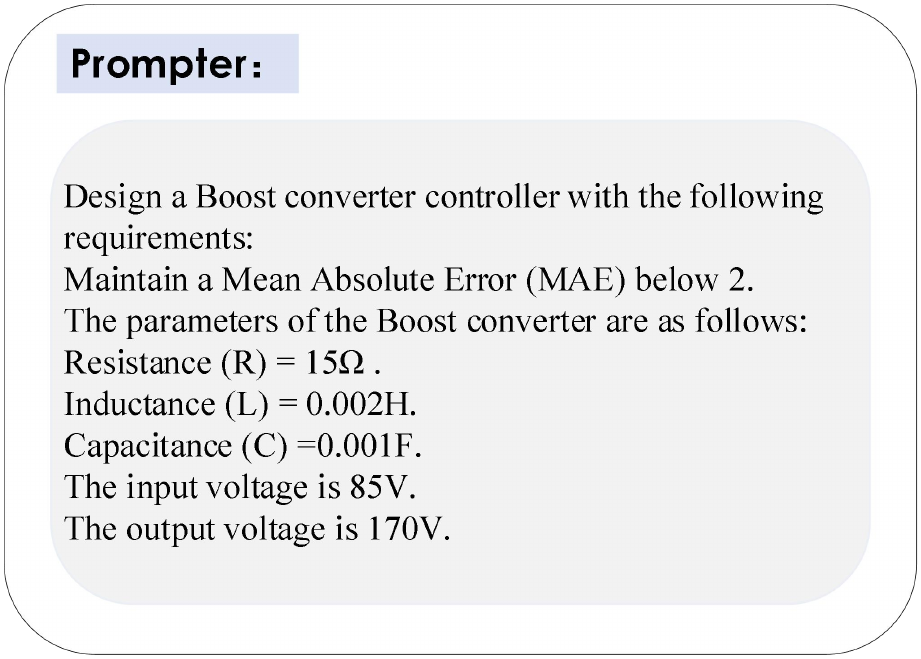}}
\caption{Designing objective oriented controller prompts.}
\label{Prompter}
\vspace{-0.8em}
\end{figure}

 In the process of designing objective oriented controller prompts, we need to specify the type of power electronic device for which the controller is being designed, the key parameters of the device, and the objectives that need to be achieved by the controller design. This will help in constructing the overall design of the controller. The design prompts for a DC-DC boost controller is shown in  Figure \ref{Prompter}.

\subsection{Tools Setup}

In this experiment, we have implemented several tools to facilitate the design and optimization of our DC-DC Boost controller. These tools are designed to help us implement the controller algorithm, optimize the controller, and design the model of the DC-DC Boost circuit.
\subsubsection{Controller Algorithm Tool}
We have implemented a Proportional-Integral-Derivative (PID) control algorithm in our controller algorithm tool. The PID controller is a widely used control algorithm in the industry due to its simplicity and effectiveness. It can control the system based on the current error, accumulated past errors, and prediction of future errors.
\subsubsection{Optimization Tool}
Our optimization tool employs the PSO algorithm. PSO is a computational method that optimizes a problem by iteratively trying to improve a candidate solution with regard to a given measure of quality, such as minimizing the error of the controller.
\subsubsection{Model Tool}
For the model design, we have implemented a DC-DC Boost circuit model. This model helps us to simulate the behavior of the DC-DC Boost converter and test our controller under different conditions. This tool is crucial for the design and testing of our controller before implementing it in a real-world scenario.

\subsection{Results}
As shown in Figure 4 and 5, the results of our experiment demonstrate the effectiveness of our multi-agent Framework approach for task-oriented controller design of a DC-DC converter. The user only needs to specify the task requirements prompt, and the LLM based agents can provide the controller design, controller parameters, and control performance by multi agent collaboration. The experiment demonstrates the effectiveness of a multi-agent mining approach for objective-oriented design of a DC-DC converter. By incorporating a PID control algorithm, Particle Swarm Optimization, and a DC-DC Boost circuit model into the controller design process, it is able to successfully design a controller that meets the objective requirements specified by the user. The results, as shown in Figure 5, confirm that the intelligent agent can provide the controller design, controller parameters, and control performance indicators based on the user's objective requirements. This approach not only simplifies the design process but also increases the efficiency and accuracy of the controller design.
\begin{figure}[htbp]
\vspace{-0.7em}
\centerline{\includegraphics[width=1.0\linewidth]{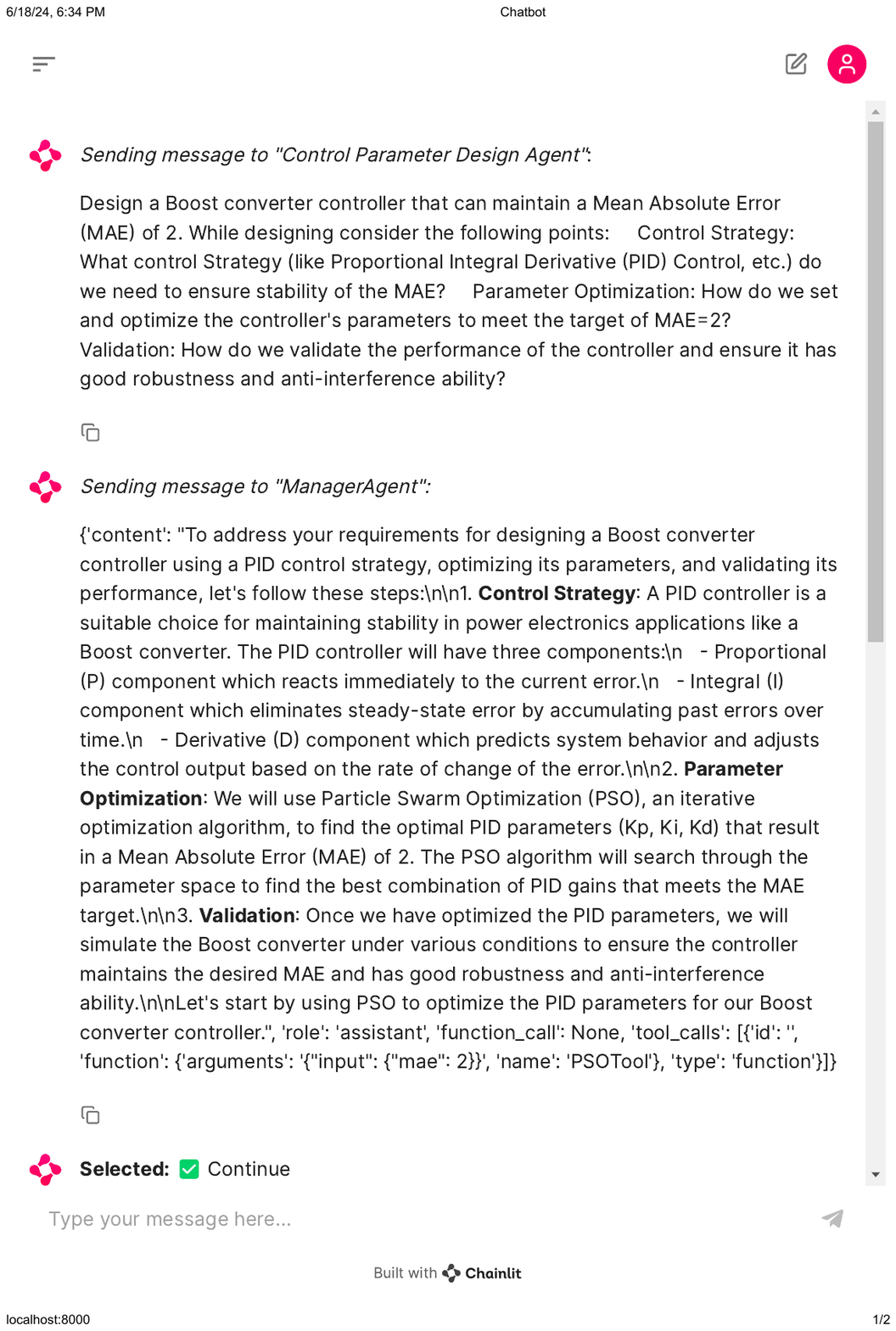}}
\caption{Web interfaces of the multi-agent framework for task-oriented controller design}
\label{Chatbot}
\vspace{-0.2em} 
\end{figure}
\begin{figure}[htbp]
\vspace{-0.7em}
\centerline{\includegraphics[width=1.0\linewidth]{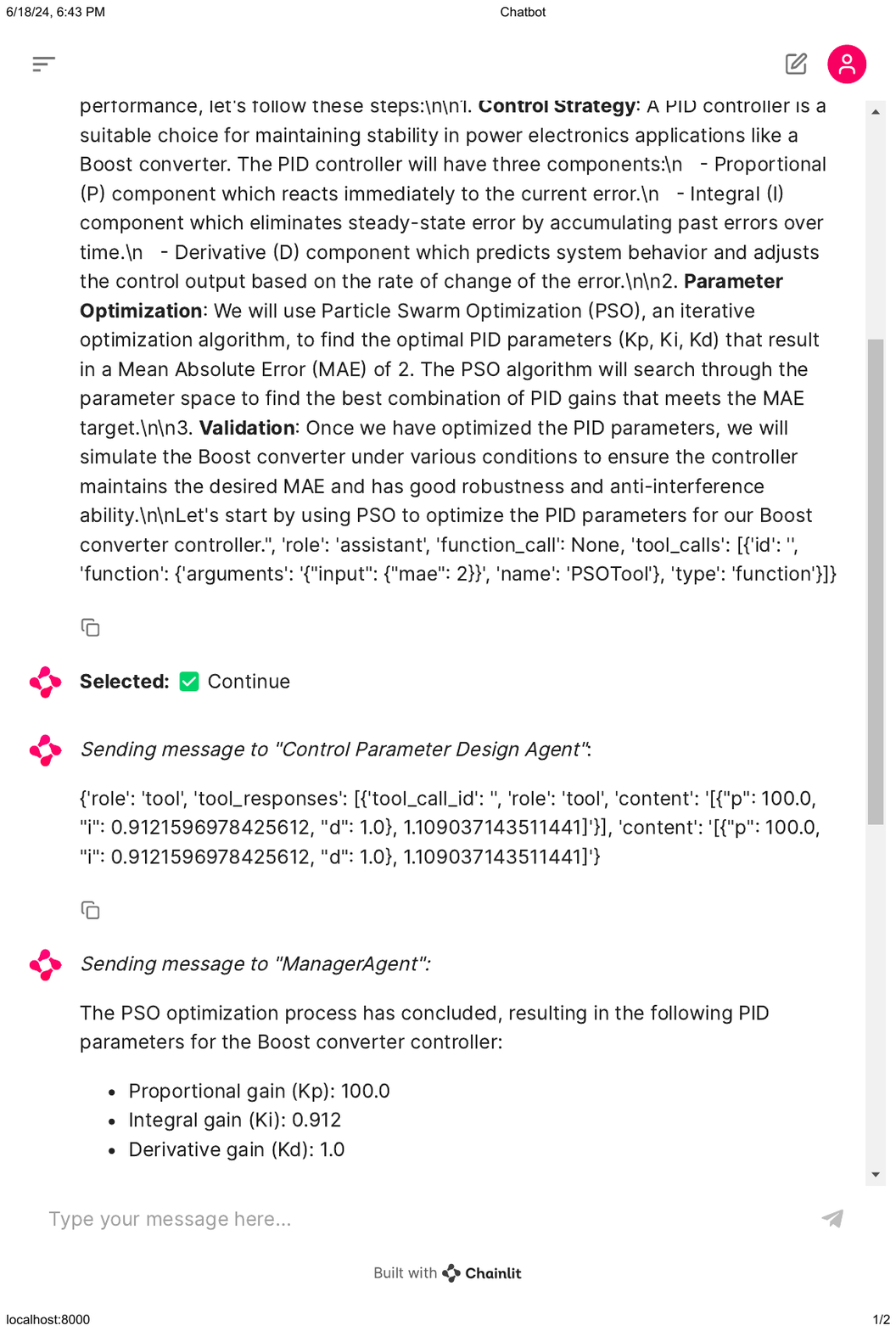}}
\caption{Results of DC-DC boost converter controller Design}
\label{5}
\vspace{-0.2em} 
\end{figure}
\section{Conclusion}
In this work, we propose a novel Large Language Model (LLM)-based multi-agent collaboration framework for objective-oriented controller design in power electronics. Unlike traditional controller design that focuses on meeting design indicators, our approach utilizes multi-agent collaboration to realize the target tasks of controller design, thereby achieving automated design of power electronic controllers.
We have initially implemented this framework through Model design, Controller algorithm, Optimization tools. Finally, we applied it to the objective-oriented design of a DC-DC boost converter controller, which verified the feasibility of our framework.
This new approach opens a promising direction for the controller design and optimization of power electronics. It will significantly contribute to the advancement of controller design in the field of power electronics.

\bibliographystyle{ieeetr}
\bibliography{ref}

\vspace{12pt}

\end{document}